# Electronic Transport Properties of Pentacene Single Crystals upon Exposure to Air

*Oana D. Jurchescu, Jacob Baas, and Thomas T.M. Palstra*[*]

Solid State Chemistry Laboratory, Materials Science Centre,

University of Groningen

Nijenborgh 4, 9747 AG Groningen, The Netherlands

**Abstract:**

We report the effect of air exposure on the electronic properties of pentacene single crystals. Air can diffuse reversibly in and out of the crystals and controls the physical properties. We discern two competing mechanisms that modulate the electronic transport. The presence of oxygen increases the hole conduction, as *in dark* four $O_2$ molecules introduce one charge carrier. This effect is enhanced by the presence of visible light. Contrarily, water, present in ambient air, is incorporated in the crystal lattice and forms trapping sites for injected charges.

[*]Corresponding author: t.t.m.palstra@rug.nl

Pentacene is a good material for electronic devices like field effect transistors ([1-8]). However, the results obtained by different research groups are not always consistent. This can be attributed in part to the fact that the performance of the devices depends critically on the environmental conditions [9-11]. In many cases these conditions are not explicitly stated, so a comparison between them is often not appropriate.

We report the electronic properties of pentacene single crystals under controlled conditions of pressure, gas composition, illumination, and in time. We provide experimental evidence that the influence of air on the charge carrier conduction consists of two separate, distinguishable contributions. Air diffuses into pentacene and can be completely removed by re-evacuation. Thus, no irreversible chemical reactions, like oxidation of pentacene to pentacenequinone, can be detected during the diffusion process. We demonstrate that oxygen absorption is responsible for an increase in the conductivity, and its effect is enhanced by the presence of visible light. In contrast, water present in ambient air will form trapping sites for the injected charges. This will reduce the number of free charge carriers, and thus the value of the conductivity. These opposite effects may account for the contradicting reports on the influence of ambient conditions.

Pentacene single crystals with a high degree of purity were grown using the vapour transport technique [12]. The growth was preceded by the purification of the starting material using a two-step vacuum sublimation process under a temperature gradient described elsewhere [8]. We performed thermo-gravimetric analysis (TGA) on pentacene single crystals at room temperature to determine the changes in mass. The increase in weight at normal pressure in the presence of different gas flows (air, $O_2$, $N_2$, and Ar) were measured using a SDT 2960 from Thermal Analysis Instruments.

Silver epoxy contacts were painted on an unexposed crystal. The electrical measurements were performed in a home built micromanipulator probe station, connected to a turbo pump and to gas supplies. The system is equipped with a glass window for optical inspection. This window is covered for the experiments performed *in dark* (Fig. 2). Otherwise, the samples are exposed to ambient fluorescent light (Fig. 3), with no additional illumination. A Hewlett Packard 4155B Semiconductor Parameters Analyser was used to perform electrical measurements. The typical hold and delay time for these scans were 10 ms in order to avoid charging of the material. All measurements were performed at room temperature.

Fig. 1 shows the evolution in time of the current density J versus electric field E characteristics for single crystal pentacene *in dark*, after venting the system with dry and ambient air, respectively. The contacts were observed to be ohmic. In the ohmic regime ($J \propto E$), no marked changes in the conductivity are induced by the presence of oxygen or water. Apparently, no free charge carriers are introduced by $O_2$ absorption. However, the space-charge-limited current regime SCLC ($J \propto E^2$) is influenced by exposure to air. Here, J increases upon exposure to pure air, whereas J decreases upon exposure to ambient air. We interpret this behaviour that the induced carriers are located in shallow trapped states and are only released beyond a critical field $E_c \approx 10^4$ V/m. The quantitative analysis of the transport changes, introduced by absorbing air, are performed in the SCLC regime ($E = 3.33 \cdot 10^4$ V/m). In dry air, the number of holes introduced by the $O_2$ molecules increases in time, leading to an increase in the charge carrier density. The system will achieve the SCLC regime at smaller bias voltage. In ambient air, as more water is accumulated, more traps are introduced and the value of this critical field will increase in time.

In the inset included in Fig1 we plot the weight increase of pentacene single crystals upon exposure to 5N dry air, at room temperature and atmospheric pressure, expressed as a molar fraction of air in pentacene. The mass of the accumulated air in the crystal structure of pentacene was calculated with the assumption that the change in mass is caused only by absorption of gases. This assumption is supported by the observation that the process is completely reversible. The same trend is observed for the measurements performed in pure $O_2$, $N_2$, and Ar. However, the time constants and the equilibria are slightly different, as they are determined by the size, shape of both crystals and diffusing molecules and their interactions.

We used Fick's second law to model the diffusion of gases into pentacene single crystals. We solve this equation for the one-dimensional case, constant diffusion coefficient *D*, and an inexhaustible source. The total quantity of gas molecules in the pentacene crystal is obtained by integrating this solution over the length of the diffusion profile. The result is normalized and expressed in molar fraction of gas molecules in pentacene. An extra term needs to be added in the solution ($N_{start}$). This term reflects the fact that the measurement starts at $t=0$ with a quantity of gas that was accumulated in crystal during venting to atmospheric pressure and subsequent handling. The equation that models the diffusion of gasses in pentacene is thus:

$$N(t) = N_{start} + N_{source} \cdot \left[ d\left(1 - erf\left(\frac{d}{2\sqrt{Dt}}\right)\right) + \frac{2}{\sqrt{\pi}} \cdot \sqrt{Dt} \cdot \left(1 - \exp\left(-\frac{d^2}{4Dt}\right)\right)\right] \qquad (1)$$

where *N(t)* is the molar fraction of gas at time *t*, $N_{source}$ is imposed by the gas flow, *d* is the length of the crystal in the direction in which the 1D diffusion occurs, *D* is the diffusion coefficient of gas in pentacene single crystals and $N_{start}$ is the value of the molar fraction accumulated before the TGA measurement. For the case of dry air (inset Fig.1), fitting the diffusion curve yields d = 17.5μm, D = $1.8\cdot10^{-10}$ $cm^2/s$ and $N_{start} = -1.37\cdot10^{-3}$ molecules air/molecules pentacene. Pentacene has a layered structure with a herringbone arrangement within the layers. The microscopic mechanism of absorption in pentacene crystals is not known. Our results show that the diffusion length is similar to the thickness of the crystal. Therefore, the diffusion proceeds via the largest surface area, perpendicular to the layers. This suggests that the oxygen molecules may not accumulate between the van der Waals bonded layers, but within the layers. A possibility is that the oxygen forms a *reversible* 6,13-adduct, which is the most reactive position of pentacene. This adduct could yield *irreversibly*, under conditions not explored here, via the formation of the hydroquinone, the 6,13-pentacene-quinone, which is conventionally the dominant impurity species [8].

We investigate the influence of the absorbed gas molecules on the evolution of the electrical current. Prior to the experiments, the samples were outgassed in vacuum (p = $5\cdot10^{-7}$ mbar) in the measurement chamber for at least 100 min, to remove the incorporated gas from the solid. The measurements can be reproduced after re-evacuation, as the gases diffuse reversibly in and out of the crystal. In order to study the influence of external factors on the electrical properties, the pentacene single crystals were exposed to both dry (purity 5N) and ambient air using a leak-valve connected to the measurement chamber. For the experiments performed *in dark*, we measured both while outgassing and venting the chamber to ensure that the changes originate from diffusion and not from charging. The evolution of the electrical properties in time was recorded immediately after venting the system to atmospheric pressure. The initial time, $t = 0$, is taken when pressure equals atmospheric pressure.

The four panels in Fig. 2 present the evolution of the electrical current at fixed bias versus pressure (2a, 2c) and time (2b, 2d) at atmospheric pressure *in dark*. The effect of oxygen exposure is studied when venting with dry air (2a, 2b). As the pressure increases (Fig. 2a), more oxygen will diffuse into the crystal, and the conductivity increases. Once the system is at atmospheric pressure, the current continues to increase (Fig. 2b) because the diffusion still continues, and its rate determines the response time. A closer inspection of the inset in Fig. 1, together with Fig 2b reveals a remarkable consistency between the absorption of gas and the electronic response. The time evolution of the electrical current mirrors that of the oxygen absorption. When venting with ambient air, both water and oxygen are introduced into the chamber (Fig. 2c, 2d). We noticed that in ambient air, the effect of oxygen exposure is counteracted by the exposure to water. Analysing the magnitude of the induced changes in the current by combining oxygen and water, one can distinguish two pressure regimes (Fig. 2c). At low pressure ($p < 10^{-2}$ mbar), the conduction is independent of air pressure. This leads us to conclude that the two effects are opposite and of the same magnitude. At $p = 10^{-2}$mbar, a sudden drop in the current is observed. Apparently, for $p > 10^{-2}$mbar the introduction of scattering sites exceeds the creation of holes, and this has a major effect on the conduction. This is reflected also in the results at atmospheric pressure, where diffusion of dry air causes a current increase of 25% (Fig. 2b), whereas ambient air diffusion cause a drop of 45% (Fig. 2d).

Fig. 3 shows the evolution of the current under the same conditions, but now in the presence of *light*. The effect of oxygen diffusion is stronger than *in dark* (Fig. 3a, 3b). This is reflected in the increase of the current, which is a factor 2.4 larger in dry air than in ambient air. As the effect of water is independent of the presence of *light*, the total influence of ambient air on the crystal properties will be different in the presence of *light*. Consistent with the measurements *in dark*, we find two pressure regimes. At low pressure ($p < 10^{-2}$ mbar), when only an insignificant amount of water is present, the oxygen doping is the dominant factor of the conduction mechanism. At $p = 10^{-2}$ mbar, when water diffuses considerably into the pentacene crystal, the effect of oxygen is decreased, and therefore a change in slope is observed (Fig. 3c). However, the presence of oxygen remains the most important factor, as the value of the current increases with pressure. After venting with ambient air, the doping caused by oxygen, activated by illumination, gives a stronger effect than the trapping of charges by water. The hysteresis in figures 2c and 3c is very small. This indicates that the variation in the values of the current is only caused by the diffusion of gases and not by charging.

We relate the changes that we observe in the electrical properties with the diffusion of gases into the crystal. Our model includes two competing mechanisms associated with the two different chemical species that influence the conduction, affected by external factors such as light. On the one hand, water absorbed by exposure to ambient air [13], will create new trapping sites for the charges that are injected from the electrodes [14]. This will decrease the value of the electrical conductivity. This effect is independent of the exposure to *light*. The polar nature of the $H_2O$ molecules causes interactions with the injected charges and increases the energetic disorder [15]. This will lead to a decrease of the conduction. Oxygen present in air will be incorporated in the crystal structure, as shown in the inset in fig. 1. Because of its electronegativity, it will attract electrons from the pentacene molecules and this process will generate holes. The density of charge carriers responsible for the conduction will thus increase, and accordingly the value of the current will increase. This process is enhanced by the presence of *light*. The measurements performed in pure $N_2$ and Ar under identical conditions show insignificant changes in the electrical properties, even though TGA indicates a similar amount of absorption.

The value of the hole current is affected by the number of gas molecules present in the crystal and thus varies in time: $J_p(t) = e\mu_p p(t)E$, where $e$ is the elementary charge and $\mu_p$ is the mobility of holes in pentacene. The term $p(t)$ is the density of charge carriers and consists of several terms: $p(t) = p_{inj} + p_{dop} - p_{traps}$, where $p_{inj}$ represents the hole density injected from the contacts, and $p_{dop}$ is the hole density induced by exposure to oxygen, which is a fraction of the number of air molecules absorbed in the solid, $p_{dop} = 0.2 \cdot \eta_{oxygen} \cdot N(t)$. The oxygen efficiency $\eta_{oxygen}$ is thus defined as the inverse of the number of oxygen molecules required to generate one hole. The term $p_{traps} = p_{def} + p_{imp} + p_{water}$ is given by the density of traps in the band gap. In vacuum a fraction of the injected charges are trapped by crystal defects, $p_{def}$, and impurities, $p_{imp}$. In ambient air, an extra term is added in the trapping sites due to the presence of water molecules, $p_{water}(t) = \beta N(t)$, which is a fraction of total number of the air molecules $N(t)$ accumulated in the crystal. Combining the measurements performed in pure air (Fig. 2b, 3b), with *N(t)* from Fig. 1(inset), we can determine the oxygen efficiency $\eta_{oxygen}$. This number is constant and independent of time with $\eta_{oxygen} \approx 0.25$ in dark and $\eta_{oxygen} \approx 0.5$ in light. This means that the efficiency of the oxygen exposure induced doping is independent of oxygen loading and it is increased by light exposure. Roughly two $O_2$ molecules are needed *in light* to create one hole, compared with four molecules *in dark*.

In ambient air, the number of trapping sites is proportional to the number of water molecules that diffuse into the crystal. We introduce the ratio:

$$\Theta_{water} = \frac{p_{inj} - p_{water}}{p_{inj}} \tag{2}$$

that quantifies the fraction of untrapped charge carriers: $\Theta_{water} = 1$ in vacuum. After exposure to ambient air, the value of $\Theta_{water}$ is reduced. $\Theta_{water}$ is calculated by substracting Fig. 2b (3b) from Fig. 2d (3d), using appropriate geometrical factors. The effects of water *in dark* and *light* have similar magnitudes, and they follow the diffusion of water, introduced from ambient air.

The changes of the conductivity with time and pressure are the electronic response to the diffusion of gases in pentacene crystals. They are fully reversible. The p-doping by molecular oxygen increases the concentration of charge carriers in shallow traps. Water molecules form new electronic states in the gap that trap the injected charges. It should be noticed that experiments under UHV conditions will not be influenced by exposure to an ambient atmosphere, as the absorbed gas can be fully removed by evacuation [16]. We expect that these effects are even larger in thin films, in which the presence of grain boundaries facilitates the diffusion.

In conclusion, we have investigated the intercalation of gases in pentacene single crystals and its effect on the electronic properties. We differentiate between the exposure to oxygen and water. Absorbed oxygen enhances the conduction by introducing holes near the valance band. Absorbed water molecules create new defect states, which trap the injected charges. This insight in the relationship between the macroscopic diffusion of gas in organic conductors and the effect on the physical properties may lead to an improved control of the device performance.

**Acknowledgements:**

The authors like to thank Petra Rudolf, Norbert Koch, Ria Broer and Paul de Boeij for stimulating discussions and Paul Blom for careful reviewing of the manuscript. This work is part of the MSC$^{plus}$ program.

## FIGURE CAPTIONS

### FIGURE 1:

Current density J vs. electric field E for pentacene single crystal at room temperature in vacuum and after exposure to dry air (increase in J) and ambient air (decrease in J). The different color of the curves represent different exposure times. In pure air: — 250 min, — 200 min, — 150 min, — 100 min, — 50 min, — 10 min, — 0 min, — in vacuum. In ambient air — 0 min, — 10 min, — 20 min, — 30 min, — 50 min, — 100 min, — 150 min, — 200 min, — 250 min.

The inset represents the evolution in time of the molar fraction of air in pentacene upon exposure of single crystals to dry air at room temperature and the fit with a 1D diffusion model. This is a lower limit of the molar fraction, as it is assumed that the gas molecules diffuse homogenously over the crystal. It is possible that the gas accumulates in the first layers, only.

### FIGURE 2:

Pressure (2a, 2c) and time (2b, 2d) dependence of the current at fixed bias voltage for single crystal pentacene in dark. The effect of dry air is shown in figures 2a and 2b (■) and of ambient air in 2c and 2d (□).

### FIGURE 3:

Dependence of the current at fixed bias voltage for single crystal pentacene of pressure (3a, 3c) and time at normal pressure (3b, 3d) exposed ambient florescent light. The effect of dry air is shown in figures 3a and 3b (■) and of ambient air in 3c and 3d (□).

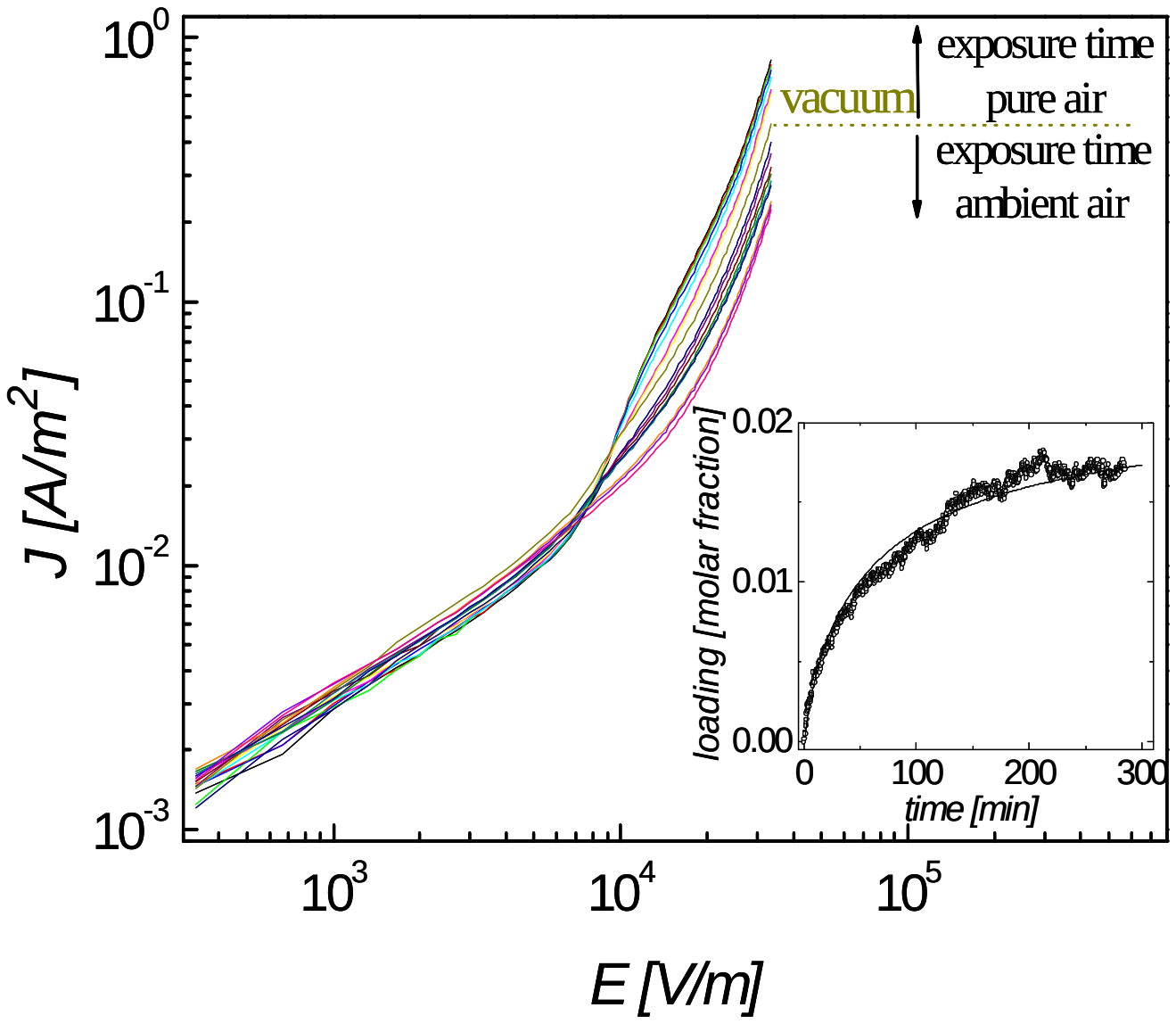

Fig1. Jurchescu et al.

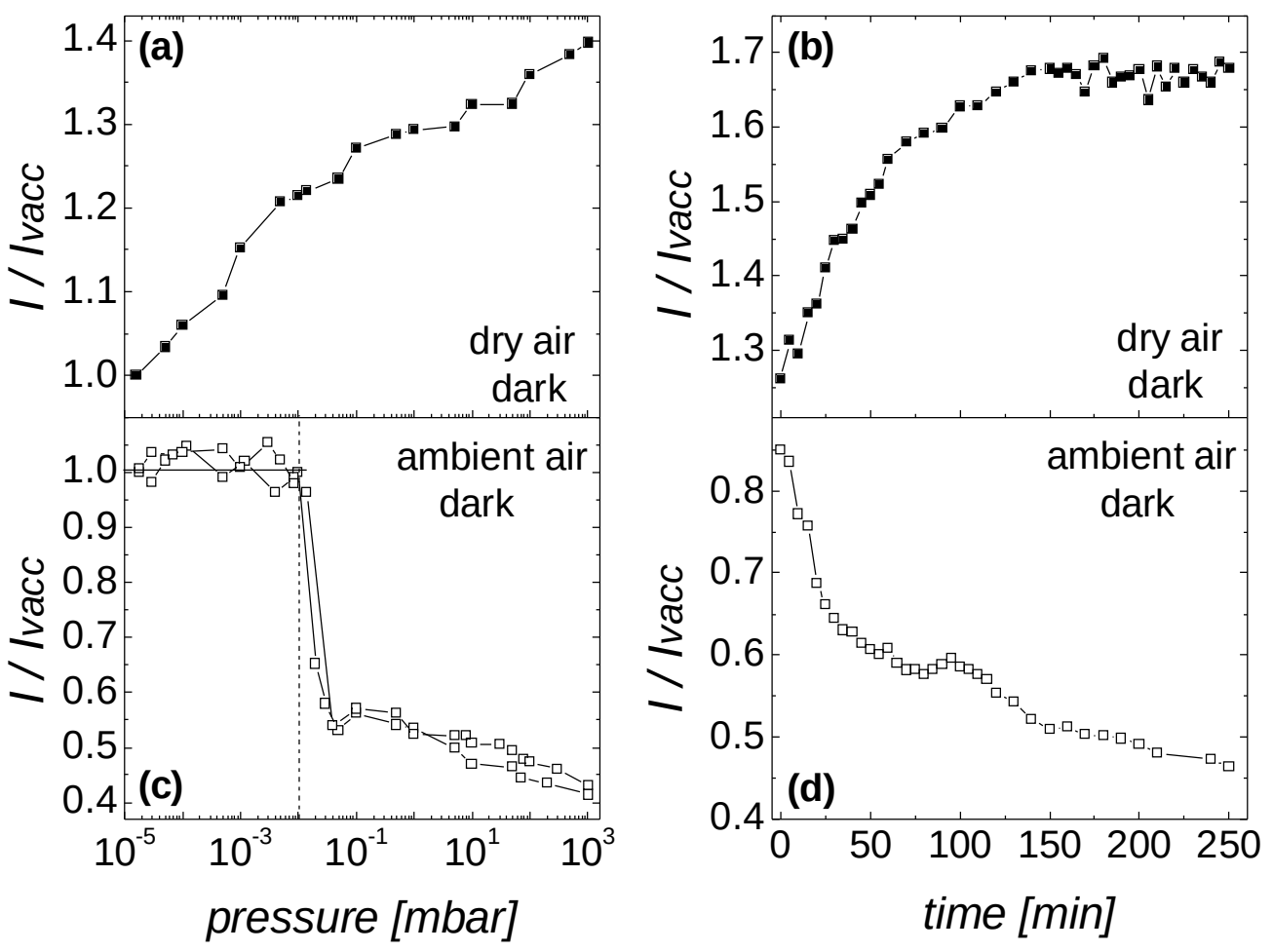

Fig 2. Jurchescu et al.

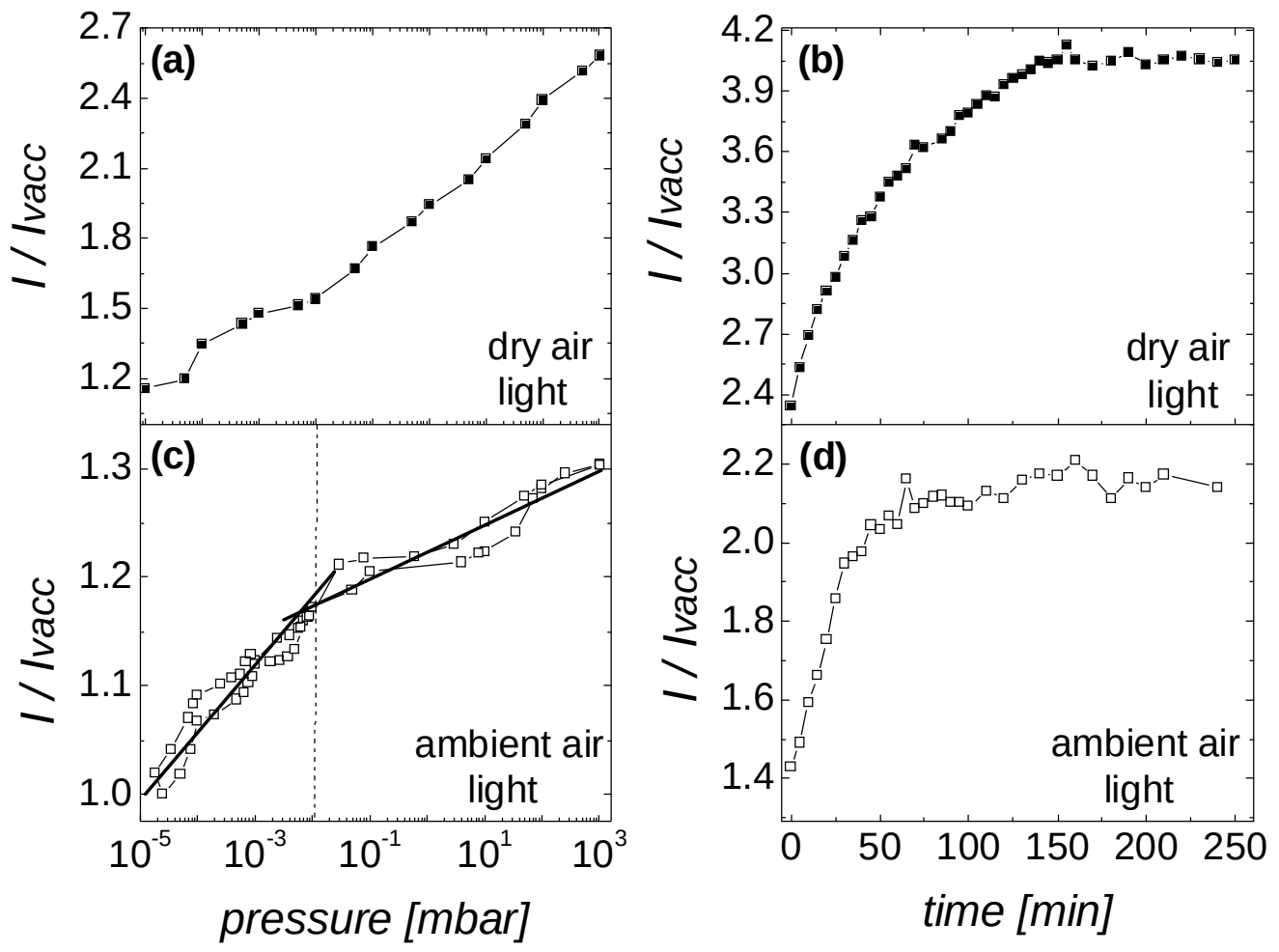

Fig 3: Jurchescu et al.